\documentclass[twocolumn,showpacs,preprintnumbers,amsmath,amssymb]{revtex4}

\usepackage[latin1]{inputenc}
\usepackage{graphicx}
\usepackage{indentfirst}
\usepackage{amsmath, amsthm, amssymb} 
\input epsf
\usepackage{float}
\usepackage[colorlinks=true, a4paper=true, pdfstartview=FitV, linkcolor=blue, citecolor=blue, urlcolor=blue]{hyperref}

\begin{document}  

\pagestyle{headings}

\title{Satellite rings and normal modes in rotating clouds of ultra cold atoms} 

\author{J.D. Rodrigues}
\author{J.T. Mendon\c ca}
\affiliation{IPFN, Instituto Superior T'{e}cnico, Avenida Rovisco Pais 1, 1049-001 Lisboa, Portugal}

\author{J.A. Rodrigues}
\affiliation{FCT, Universidade do Algarve, Campus de Gambelas, Faro, Portugal}


\begin{abstract}
\begin{footnotesize}

The multiple scattering of light in a gas of ultra cold atoms is responsible for many exciting features observed in magneto-optical traps including the collective behaviour forced by a Coulomb like potential. This field also induces plasma like phenomena in the cloud which allows the treatment of the system as a one component trapped plasma. With a fluid description and casting the thermodynamical behaviour in the form of a polytropic equation of state we investigate the equilibrium profiles of rotating clouds and its dependence on the experiment characteristics. Numerical solutions predict the formation of stable orbital modes both in rotating and non rotating clouds. We also investigate the normal modes on such rotating systems. The results obtained allow for the measurement of the equation of state for the cold gas.

\end{footnotesize}
\end{abstract}

\maketitle

\section{Introduction}

The celebrated laser cooling processes and magneto-optical traps (MOT)  \cite{MOT,metcalf}  have allowed the study of many exciting topics in atomic physics. Among them, much interest was devoted to the study of Bose-Einstein condensates  \cite{BEC01,BEC02}, which had a profound impart in our understanding of condensed matter. Apart from this, a new trend has recently begun that rekindled the interest on basic properties of MOT physics. This is related to the increasing number of astrophysical phenomena that we can simulate and study using ultra cold atomic clouds. In particular, we can refer to a new mechanism associated with the laser cooling process, which can lead to the formation of static and oscillating photon bubbles inside the gas \cite{bubbles}. Photon bubbles have been considered in the astrophysical context \cite{b1,b2,b3} where huge photon densities are required to have any significant impact on high energy particles. 
\par
Moreover, Kaiser \textit{et al} \cite{random} were recently able to achieve random lasing in a cloud of ultra cold atoms under laboratory conditions.  A similar effect was first seen decades ago in stellar clouds \cite{r2,r3} and in some planetary atmosphere \cite{r4}, when random lasing was first proposed to explain why certain specific emission lines in the stellar gas are more intense than theoretically expected \cite{r3, r6}.
\par
Finally we refer to a recent work by Ter\c{c}as \textit{et al} \cite{hugo}, where the hydrodynamic equilibrium and normal modes of cold atomic traps are investigated, combining the effects of multiple photon scattering with the thermal fluctuations inside the system, cast in the form of a polytropic equation of state. This analysis results in a generalized Lane-Emden equation, describing the equilibrium density profiles of the atomic cloud, similar to that describing astrophysical fluids \cite{lane}.
\par
The process of multiple scattering of light, which typically becames significant for a number of atoms above $N \simeq 10^{5}$, is responsible for the rich and complex behaviour of ultra cold atomic vapours. This mechanism has been described, since the early stages of MOTs, as the principal limitation for the compressibility of the cloud \cite{mot1,mot2}. In this regime, the atoms in the cloud are strongly correlated due to the presence of a Coulomb type long-range interaction \cite{mot3}, and the description of the system as a one component plasma becomes feasible and very fruitful \cite{p1,p3,p4}.
\par
In the present paper, we present a significant extension of the previously mentioned work by Ter\c{c}as \textit{et al} \cite{hugo}, to include the case of a rotating cloud of ultra cold atoms. In section II, we begin with the derivation of the equilibrium density profiles of such systems. Numerical analysis predict the existence of a special set of solutions containing satellite rings. Such satellite rings have indeed been observed in rotating clouds since the early nineties \cite{mot2, 1ring}, although their nature was not completely elucidated. In section III, we will investigate the stability of such systems, by computing the normal oscillation modes that can be excited in the cloud. Finally, in section IV, we discuss the  validity of our model and state some conclusions.

\section{Polytropic equilibrium}

\par
The starting point to compute the equilibrium profiles of the ultra cold cloud of atoms confined in a MOT corresponds to the setting of the fluid equations, where the collective force due to multiple scattering of photons by the atoms is included:
\begin{equation}
\frac{\partial n}{\partial t} + \boldsymbol{\nabla} \cdot (n \boldsymbol{v} ) = 0 \, , 
\end{equation}
\begin{equation}
\frac{\partial \boldsymbol{v}}{\partial t} + \boldsymbol{v} \cdot \boldsymbol{\nabla} \boldsymbol{v} = -\frac{ \boldsymbol{\nabla} P}{m n} + \frac{\boldsymbol{F}_{T}}{m} \, .
\end{equation}
The fluid description of the system requires some relation between the pressure $P$ and the atom density $n$, and for that purpose we assume the existence of a generic polytropic equation of state for the MOT, of the form
\begin{equation}
P(r) = C_{\gamma} n(r)^{\gamma} \, .
\end{equation}
As we are dealing with rotating clouds, the total force acting on an fluid element is $\boldsymbol{F}_{T} = \boldsymbol{F}_{MOT} + \boldsymbol{F}_{c} + \boldsymbol{F}_{r}$, with $\boldsymbol{F}_{MOT} \simeq - \alpha \boldsymbol{v} - \kappa \boldsymbol{r}$, and $\boldsymbol{F}_{c}$ the collective force determined by 
\begin{equation} \label{eq:lap1}
\boldsymbol{\nabla} \cdot \boldsymbol{F}_{c} = Qn \, .
\end{equation}
Notice that in the expression for the collective force $\boldsymbol{F}_{c}$, the quantity $Q = (\sigma_{R} - \sigma_{L})\sigma_{L} I_{0} / c > 0$ represents the square of an effective atomic charge \cite{mot3}, where $c$ is the speed of light and $I_{0}$ the total intensity of the six cooling laser beams. The terms $\sigma_{R}$ and $\sigma_{L}$ represent the emission and absorption cross sections, respectively \cite{mot2}. The force $\boldsymbol{F}_{MOT}$ includes the Doppler cooling force, with an equivalent damping coefficient $\alpha$, and the trapping force, with an equivalent spring constant $\kappa = m \omega_{0}^{2}$. The difference with respect to the non rotating case is the presence of a new force term $\boldsymbol{F}_{r}$. Rotation in the system can easily be achieved by a slight misalignment in four of the six laser beams, and can be  described by  \cite{mot2}
\begin{equation}
\boldsymbol{F}_{r} = \kappa' r \boldsymbol{e}_{z} \times \boldsymbol{e}_{r} = \kappa' r \boldsymbol{e}_{\phi} \, .
\end{equation}
For this reason, we will consider from now on a cylindrically symmetric system.
Assuming equilibrium conditions, $\partial / \partial t = 0$ and $\dot{r} = 0$, and a-dimensioning the system as
\begin{equation}
\theta (r) \equiv \left( \frac{n(r)}{n(0)} \right)^{\gamma - 1} \, ,
\end{equation}
and $\xi \equiv r / R$ with 
\begin{equation}
R = \left( \frac{P(0)}{3 m \omega_{0}^{2} n(0)} \right)^{1/2} \, ,
\end{equation}
we then get, in a dimensionless form
\begin{equation} \label{eq:Lane0}
\frac{\gamma}{\gamma -1} \frac{1}{\xi} \frac{\partial}{\partial \xi} \left( \xi \frac{\partial \theta}{\partial \xi} \right) + (1 - \beta^{2}) - \Omega_{p}'^{2} \theta^{1/ (\gamma - 1)} = 0 \, ,
\end{equation}
with 
\begin{equation}
\Omega_{p}'^{2} = \frac{Q n(0)}{3m \omega_{0}^{2}} \equiv \frac{\omega_{p}^{2}}{3 \omega_{0}^{2}} \, ,
\end{equation}
the effective plasma frequency, and $\beta^{2} = 2 \left( \dot{\phi} / \omega_{0} \right)^{2} \geq 0$, where $\dot{\phi}  = \kappa' / \alpha$ is the angular velocity of the fluid element. The parameter $\beta$ gives the ratio between the rotation angular frequency and the frequency associated with the confinement trap, $\omega_{0}$. The first is equivalent to an expansion force, and the second one to a contraction force, whereby this constant will be related with the stability of the cloud. In particular, the system will become unstable for $\beta^{2} > 1$, as it will become clear in section \textbf{III}, when we derive the frequency of the allowed oscillation modes. For non rotating clouds we would simply have $\beta = 0$. Eq. (\ref{eq:Lane0}) therefore depends on the rotation state of the system. We can now realize that, by redefining the parameter $R$ as
\begin{equation}
R = \left( \frac{P(0)}{3 m \omega_{0}^{2} n(0)} \right)^{1/2} \rightarrow   \left( \frac{P(0)}{3 m \omega_{0}^{2} (1- \beta^{2}) n(0)} \right)^{1/2} \, ,
\end{equation}
and introducing a redefined plasma frequency as 
\begin{equation}
\Omega_{p}^{2} = \frac{\Omega_{p}'^{2}}{1 - \beta^{2} } \, ,
\end{equation}
we get
\begin{equation} \label{eq:Lane}
\frac{\gamma}{\gamma -1} \frac{1}{\xi} \frac{\partial}{\partial \xi} \left( \xi \frac{\partial \theta}{\partial \xi} \right) + 1 - \Omega_{p}^{2} \theta^{1/(\gamma - 1)} = 0 \, .
\end{equation}
This new equation becomes independent of rotation parameter, which is incorporated now inside the definitions of $R$ and $\Omega_{p}^{2}$. It  remarkably implies that rotating and non rotating systems share the some kind of solutions for the equilibrium density profiles, differing only by a scale factor $(1 - \beta^{2})^{1/2}$.  Eq. (\ref{eq:Lane}) can be seen a generalized Lane-Emden equation \cite{lane}, which was first derived for astrophysical fluids with gravitational confinement. The corresponding density profiles will only depend on the polytropic exponent $\gamma$, and on the constant $\Omega_{p}^{2}$, which accounts for the three forces in play. The latter corresponds to the {\sl ratio} of the collective force, determined by $\omega_{p}^{2}$, and the confining force ($\omega_{0}^{2}$), {\sl subtracted} by the centripetal force due to rotation ($\beta^{2}$). The confining force is then directly contour-posed by the centripetal one. This ratio is sufficient to determine the density profiles. 

With this interpretation of $\Omega_{p}^{2}$ in mind, we can easily realize that a stable cloud of atoms can only exist if $\Omega_{p}^{2} < 1$ which implies an overall attractive force greater than the repulsive one (multiple scattering). Since we are interested in associating systems with different rotation states, we realize that degenerate density solutions only exist if $\Omega_{p}^{2} = \Omega_{p}'^{2}/(1 - \beta^{2}) = \eta$, with $\eta<1$ for a stable solution. Each of these sets of solutions (for each value of the constant $\eta$) differ only by the scale factor $(1-\beta^{2})^{1/2}$. Let us now compute some analytical solutions of eq. (\ref{eq:Lane}), for some important limiting cases.

\subsection{Temperature limited regime}

 For traps with a small number of atoms, typically $N < 10^{5}$  multiple scattering effects can be neglected, which allows us to take the limit $\Omega_{p}^{2} \rightarrow 0$. Eq. (\ref{eq:Lane}) then reduces to
\begin{equation}
\frac{\gamma}{\gamma - 1} \frac{1}{\xi} \frac{\partial}{\partial \xi} \left( \xi \frac{\partial \theta}{\partial \xi} \right) + 1 = 0 \, ,
\end{equation}
which takes the solution $\theta(\xi) = \left[1- \frac{\gamma - 1}{4 \gamma} \xi^{2} \right]$, or equivalently, in dimensional form 
\begin{equation} \label{eq:gama}
n(r) = n(0) \left( 1 - \frac{\gamma - 1}{4 \gamma} \frac{r^{2}}{R^{2}} \right)^{\frac{1}{\gamma - 1}} \, .
\end{equation}
The isothermal case, which corresponds to $\gamma =1$, simply yields a Gaussian profile
\begin{equation}
n(r) =  n(0) \exp \left[ -\frac{m \omega_{0}^{2} (1-\beta^{2}) r^{2}}{2 k_{B} T} \right] = n(0) \, e^{-r^{2} / 4 R^{2}} \, ,
\end{equation}
which results from taking the limit $\gamma \rightarrow 1$ in eq. (\ref{eq:gama}) \footnote{Remember the definition of the exponential function as  $ e^{x} = \lim_{n \to \infty} \left( 1 + \frac{x}{n} \right)^{n} $}.
This limit approximately describes a trap in the temperature limited regime, corresponding to a small number of particles, where the thermal effects determine the cloud dynamics and multiple scattering effects are negligible. For a larger number of atoms, as typical of present state-of-the-art traps, this no longer applies.

\subsection{Multiple scattering regime}

In the opposite case of a large number of trapped atoms, the collective force dominates over temperature effects, and we can take the limit $\gamma \rightarrow 0$ in eq. (\ref{eq:Lane}), giving rise to the solution $\theta ( \xi) = 1/ \Omega_{p}^{2}$ or, in dimensional form
\begin{equation} \label{eq:wb}
n(r) = \frac{3}{Q}  m \, \omega_{0}^{2} ( 1 - \beta^{2}) H (a - r) \, ,
\end{equation}
where $a$ is the cloud radius and $H$ is the Heaviside function. This is usually referred to as the \textit{water-bag} profile.

\subsection{Equilibrium orbital modes}

\begin{figure}
\centering
\includegraphics[width=1\linewidth]{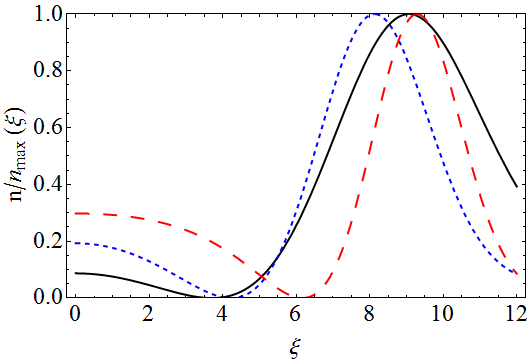}
\caption{Solutions for the generalized Lane-Emden equation, for $\gamma = 3/2$. The black line corresponds to $\Omega_p^2 = 0.2$, and for the blue and red dashed lines we have $\Omega_p^2 = 0.5$ and $\Omega_p^2 = 0.9$ respectively. The differential equations were integrated using an explicit Runge-Kutta method, of fourth order.}
\label{fig:orbital}
\end{figure}

The formation of stable orbital modes in the density profiles is one of the most interesting features observed in rotating clouds of atoms \cite{mot2}. It was widely believed that the existence of such modes was intrinsically related with the rotation of the cloud \cite{mot2, 1ring, 2ring}. The investigation presented here indicates that this is not entirely true, since rotating and non rotating clouds share the same density solutions. In fact, numerical integration of the generalized Lane-Emden equation (\ref{eq:Lane}), reveals a special set of solutions for $\gamma = 3/2$. These appear as central cores with rings of atoms around, as depicted in figure \ref{fig:orbital}.

In these solutions, there is a central core of atoms surrounded by a ring, typically with higher densities than the central region. As we increase the number and density of atoms in the system or, equivalently, higher rotation angular velocities - higher $\Omega_p^2$ - the central core density increases. This behaviour has been experimentally observed, where a single ring structure can turn into a ring with a central core, for increasing intensity of the laser beams (capturing more atoms in the trap) or by increasing the laser misalignments, making the cloud to rotate faster \cite{1ring}. In our picture this corresponds to an increase of $\Omega_p^2$. The novelty of the present analysis is the description of the system in terms of an atom density function, $n(r)$, which is the experimentally accessible quantity, making possible the direct comparison of theory and experiments. Although in references \cite{mot2, 1ring} we can find a theoretical explanation for the rings, it is based on single atom trajectories, not allowing for a detailed comparison between theory and experiments. Since we relate the atom density profiles with the equation of state of the cloud, the results presented in this paper will allow, for the first time, and using already available absorption imaging techniques, to measure the equation of state of the system. In the next section, along with the stability analysis of the solutions, we will also introduce an alternative way of measuring the equation of state, through the normal modes of oscillation in the system.

\begin{figure}
\centering
\includegraphics[width=1\linewidth]{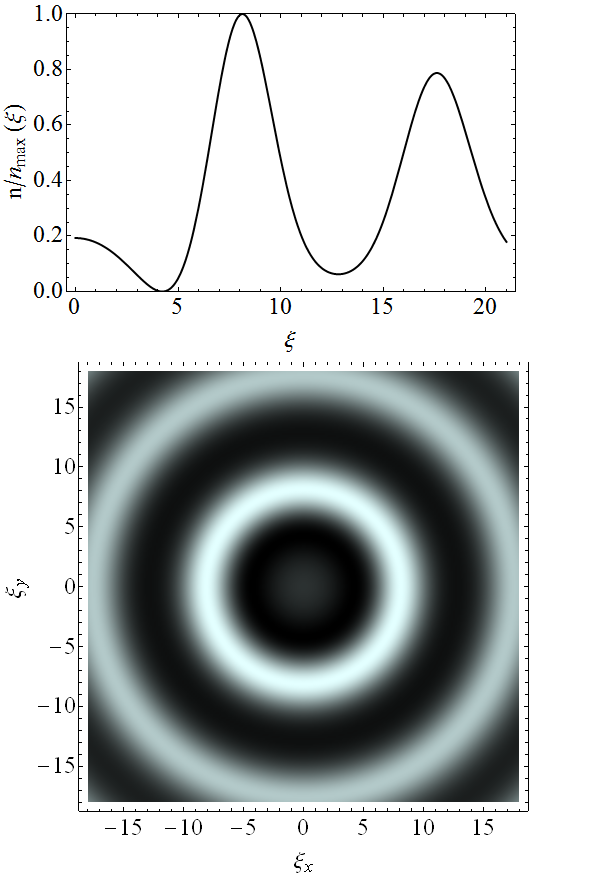}
\caption{Solution of the generalized Lane-Emden equation, for $\gamma = 3/2$ and $\Omega_p^2 = 0.5$. Two rings are displayed, which should be possible for a larger number of trapped atoms. }
\label{fig:density}
\end{figure}

In fact, the density profiles for $\gamma = 3/2$ present higher order orbital modes, with two or more distinguishable rings around the central core in spatially extended solutions. As the system is limited by the width of the laser beams and strongly depends on the number of trapped atoms, only a central core should be observed, eventually with one or two rings. Experimentally, the transition from one to two rings has been observed \cite{mot2, 2ring} when the number of trapped atoms increases. In figure \ref{fig:density} we show a numerical solution where a second ring of atoms is included. 

\section{Normal modes}

We now evaluate the nature of localized oscillations, or normal modes of the system, using a perturbative analysis on the above fluid description. It will be useful to notice that the collective force can be derived from a potential, $\phi_{c} $, such that $\boldsymbol{F}_{c} = - \boldsymbol{\nabla} \phi_{c}$. This is a Coulomb type of potential, due to the existence of an effective electric charge for the neutral atoms, and resulting from the exchange of scattered photons between nearby atoms. It can be seen as a mean-field potential associated with multiple photon scattering. We now introduce small perturbations in the equilibrium quantities, labelled with the subscript $0$,

\begin{equation}  \label{eq:oscillation}
\begin{split} 
 & n = n_{0} + \delta n \\  
 & P = P_{0} + \delta P \\
 & \boldsymbol{v} = \boldsymbol{v}_{0} + \delta \boldsymbol{v} \\
 & \phi^{c} = \phi_{0}^{c} + \delta \phi^{c} 
\end{split}
\end{equation}

with the perturbations varying as $\delta a = \bar{\delta a} e^{i\omega t}$. This prescription allows us to linearize the set of fluid equations, neglecting quadratic terms of the form $\delta a \, \delta b \sim 0$, which yields

\begin{equation} \label{eq:norm1}
\begin{split}
-\left( \omega^{2} + i \frac{\alpha}{m} \omega \right)\delta n =& \frac{\gamma C_{\gamma}}{m} \boldsymbol{\nabla} \cdot \left[ n_{0}^{\gamma - 1} (r) \boldsymbol{\nabla} \delta n \right] \\ +
 &\frac{1}{m} \boldsymbol{\nabla} \cdot \left[ n_{0}(r) \boldsymbol{\nabla} \delta \phi_{c} \right] \, ,
\end{split}
\end{equation}

\begin{equation} \label{eq:llap}
\nabla^{2} \delta \phi_{c} = -Q \delta n \, ,
\end{equation}

where we have used  $\boldsymbol{\nabla} \delta P \simeq \gamma C_{\gamma} n_{0}^{\gamma - 1} \nabla \delta n$, from the polytropic equation of state. It is worth noticing that this equation for the perturbation $\delta n$ is formally identical to that derived for non rotating clouds \cite{hugo}. This is to be expected, since we only consider radial perturbations, of the form $\delta \boldsymbol{v} = \bar{\delta v} (r) \exp (i \omega t) \boldsymbol{e}_{r}$, and $\delta n = \bar{\delta n} (r) \exp (i \omega t)$. The existence of a finite angular velocity is put in play by the equilibrium density profiles $n_{0}(r)$ from which the previous equation depends. Recall that, due to the geometry of the problem, we have used cylindrical coordinates.  As a last remark, notice that the term $i \alpha / m$ in the previous equation corresponds to a damping of the modes, which is to be expected since it results from the cooling Doppler force, which is a viscous force, $\boldsymbol{F}_{Doppler} = -\alpha \boldsymbol{v}$. In fact, along with the mode damping, this term also introduces a small correction in the mode frequency. But, as it turns out, in typical experimental conditions, the frequency associated with this term is small in comparison with the trapping frequency $\omega_{0}$, and the plasma frequency $\omega_{p}$, for what we  can neglect this correction from now on \cite{p1}. 

In order to combine equations (\ref{eq:norm1}) and (\ref{eq:llap}), we can now introduce the auxiliary quantity $\eta$, defined as $\delta n = (1 / 2 \pi r) d \eta / d r$. Making proper substitutions, the linearized equations can finally be put together, resulting in a single expression
\begin{equation} \label{eq:eq1}
\left[ \omega^{2} - \omega_{p}^{2} \frac{n_{0}(r)}{n_{0}(0)} \right] \eta (r) = -\frac{\gamma C_{\gamma}}{m} \left[ r n_{0} ^{\gamma - 1} (r) \frac{d}{d r} \left( \frac{1}{r} \frac{d \eta}{ dr} \right) \right] \, ,
\end{equation}
with $\omega_{p}^{2} = Q n(0) / m$. We have then reduced the problem of finding the normal modes of oscillation to an eigenvalue problem. Despite the linear form of the differential equation this is a non-trivial problem, and general solutions involve numerical simulations. For that reason we will only examine some limiting cases. 

\subsection{Temperature limited regime}

For small clouds, typically with $N < 10^{5}$ trapped atoms, the effects of multiple scattering can be neglected, and we can set $\omega_{p} = 0$, or equivalently $\phi_{c} = 0$ in eq. (\ref{eq:norm1}), which simplifies to
\begin{equation} \label{eq:tlr}
 \omega^{2}  \delta n + \frac{\gamma C_{\gamma}}{m} \boldsymbol{\nabla} \cdot \left[ n_{0}^{\gamma - 1} (r) \boldsymbol{\nabla} \delta n \right] = 0 \, .
\end{equation}
Replacing the equilibrium density profile $n_{0} ( r )$, given by eq. (\ref{eq:gama}), into eq. (\ref{eq:tlr}) and writing the result in terms of the adimensional variable $\xi = r/R$, as defined before, we obtain
\begin{equation} \label{eq:tlr1} 
-\omega^{2} \delta n - \frac{3}{4} (\gamma - 1) \omega_{0}^{2} \left( 1 - \beta^{2} \right) \frac{1}{\zeta} \frac{d}{d \zeta} \left[ \left( 1 - \zeta^{2} \right) \zeta \frac{d \delta n}{d \zeta} \right] = 0 \, ,
\end{equation}
where we performed another change of variables, namely $\xi = \sqrt{4 \gamma /(\gamma - 1)} \, \zeta$. For this differential equation we can try out solutions in the form of a power series
\begin{equation} \label{eq:ansatz}
\delta n  = \sum a_{\nu l} \zeta^{2\nu+l} \, ,
\end{equation}
where we can distinguish between even ($\nu$) and odd ($l$) contributions, because they lead to slightly different types of oscillations. For the lowest radial modes, $\nu=0$, the solutions correspond to surface excitations \cite{stringari}, and the even perturbations solutions, $l=0$, correspond to breathing modes. Deriving the ansatz (\ref{eq:ansatz}), and replacing in eq. (\ref{eq:tlr1}), we obtain a recursive expression for the coefficients $a_{\nu l}$, and the allowed modes of the system are 
\begin{equation}
\omega = \omega_{0} \sqrt{(1- \beta^{2}) \frac{3}{4} (\gamma - 1) (2\nu + l +2)} \, .
\end{equation}
As mentioned before, setting $\nu=0$ we get the surface modes
\begin{equation}
\omega = \omega_{0} \sqrt{(1- \beta^{2}) \frac{3}{4} (\gamma - 1) ( l +2)} \, .
\end{equation}
Breathing modes, $l=0$, are also possible even in small traps, with a spectrum given by
\begin{equation}
\omega = \omega_{0} \sqrt{(1- \beta^{2}) \frac{3}{2} (\gamma - 1) ( \nu + 1)} \, .
\end{equation}
Similar results are found in non-rotating spherical clouds \cite{hugo}. As anticipated in section \textbf{II}, the cloud becomes unstable for $\beta^{2} > 1$, when the centrifugal force due to rotation is higher than the confining force. These results, along with those from the previous section, introduce the possibility of {\sl measuring} the equation of state for the ultra cold gas, simply determined by the polytropic exponent $\gamma$. In fact, experimental techniques allow for very precise measurements of oscillation frequencies, and these results relate the polytropic exponent $\gamma$ with the modes frequencies and the angular velocity of the cloud.

\subsection{Multiple scattering regime}

In contrast with the temperature limited regime for small traps, for a large number of trapped atoms, the system is dominated by the collective effects, and we can neglect the temperature effects by setting $\gamma = 0$ in eq. (\ref{eq:eq1}), yielding
\begin{equation}
\left( \omega^{2} - \omega_{p}^{2} \frac{n_{0}(r)}{n_{0}(0)} \right) \eta (r) =0 \, .
\end{equation}
Introducing the water-bag solution for the equilibrium density profile given by eq. (\ref{eq:wb}), we obtain
\begin{equation} \label{eq:plasma_mode}
\omega = \sqrt{3} \omega_{0} \left( 1 - \beta^{2} \right)^{1/2} = \omega_{p} \, .
\end{equation}
Again, and as expected, we have an instability for $\beta^{2} > 1$, for the same reasons as before. Remember that the multiple scattering regime corresponds to the limit $\Omega_{p}^{2} \rightarrow 1$ with $\Omega_{p}^{2} = Q n (0) / 3m \omega_{0}^{2} (1 - \beta^{2})$, which implies $ \omega_{0} \sqrt{3 ( 1 - \beta^{2})} = \omega_{p}$. We then have a breathing mode in the system at the plasma frequency $\omega_{p}$. The latter result is formally analogous to that well-known in plasma physics, where uncompressional monopole oscillation of a plasma cloud takes place at the classical plasma frequency. The same solution is also found for neutral atoms with spherical symmetry \cite{hugo}. These normal modes are oscillations of the cloud, which should not be confused with the hybrid sound waves discussed in ref. \cite{p1}. In contrast with the present normal modes, which are global oscillations associated with the existence of boundaries, the hybrid sound waves are elementary excitations of the gas, which can also exist in an infinite medium. The corresponding frequencies are different, and can therefore be identified experimentally.

\begin{figure}
\centering
\includegraphics[width=1\linewidth]{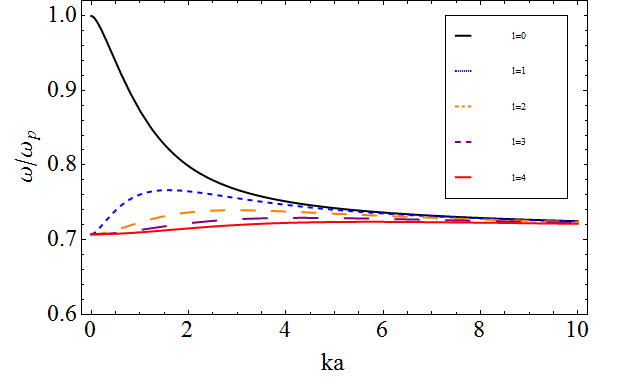}
\caption{Dispersion relation for oscillations in the multiple scattering regime, $\Omega_{p}^{2} \rightarrow 1$.}
\label{fig:dispersao}
\end{figure}

We shall, however, notice that the solution (\ref{eq:plasma_mode}) is not unique. A simple manipulation of eq. (\ref{eq:norm1}) and (\ref{eq:llap}) and introducing the equilibrium water-bag profile yields 
\begin{equation} \label{eq:eqp}
\nabla^{2} \left[ \varepsilon ( \omega ) \delta \phi_{c} \right] = 0 \, ,
\end{equation}
with $\varepsilon ( \omega ) = 1 - 3 \omega_{0}^{2}(1- \beta^{2}) / \omega^{2} = 1 - (\omega_{p}^{2} / \omega^{2})$, the effective dielectric function. This equation holds for the internal region of the cloud, while in the outside we simply have $\nabla^{2} \delta \phi_{c} = 0$, $\varepsilon(\omega) = 1$. This is quite appropriate for the water-bag density profile. The breathing mode obtained earlier can be derived from this alternative formulation by setting the particular solution to eq. (\ref{eq:eqp}) as $\varepsilon (\omega) = 0$, which implies $\omega = \sqrt{3} \omega_{0} ( 1 - \beta^{2})^{1/2} = \omega_{p}$ as before. The general solution to eq. (\ref{eq:eqp}) is given by a simple power series in spherical coordinates. In cylindrical coordinates, the solution is more complicate but it can be given in terms of the Bessel functions. We consider solutions valid in the two different regions, the inside region of the cloud, $r < a$, and the outside region, $r > a$, where $a$ stands for the cloud radius. The resulting solutions of the above Laplace equation then read \cite{bessel}
\begin{equation} \label{eq:bess}
\varepsilon(\omega) \delta \phi_{c}^{in} = A_{lk} \frac{I_{l}(kr)}{I_{l}(ka)} e^{i l \phi} e^{ikz} \, ,
\end{equation}
and
\begin{equation}
\delta \phi_{c}^{out} = B_{lk} \frac{K_{l}(kr)}{K_{l}(ka)} e^{i l \phi} e^{ikz} \, ,
\end{equation}
where $I_{l} (kr)$ and $K_{l} (kr)$ are the modified Bessel functions, and $A_{lk}$ and $B_{lk}$ are constants. We have also included the possibility of a perturbation propagating in the $z$ direction with wavevector $k$, different from the radial oscillations introduced by eq. (\ref{eq:oscillation}). These solutions have to satisfy regular continuity conditions at the surface of the cloud ($r = a$), such that
\begin{equation} \label{eq:bound1}
\varepsilon(\omega) \delta \phi_{c}^{in} (a) = \delta \phi_{c}^{out} (a) \, ,
\end{equation}
\begin{equation} \label{eq:bound2}
\left. \frac{d}{dr} \varepsilon(\omega) \delta \phi_{c}^{in} (r) \right|_{r=a} = \left. \frac{d}{dr} \delta \phi_{c}^{out} (r) \right|_{r=a} \, .
\end{equation}
The first continuity condition implies $A_{lk} = B_{lk}$, and the second condition yields the allowed modes as
\begin{equation}\label{eq:omegap}
\omega = \sqrt{3} \omega_{0} (1 - \beta^{2})^{1/2}  \left[ 1 - \frac{I_{l}'(ka) K_{l}(ka)}{I_{l}(ka) K_{l}'(ka)} \right]^{-1/2}  \, .
\end{equation}
In figure \ref{fig:dispersao},  we represent the dispersion relation given by eq. (\ref{eq:omegap}) for different values of $l$. Notice that a similar result can be found in the context of a cylindrical pore surface-plasmon modes \cite{bessel}.
We easily realize that, in the long wavelength limit, $k\rightarrow0$, we recover the breathing mode derived earlier at $\omega_{B} = \sqrt{3} \omega_{0} (1-\beta^{2})^{1/2} = \omega_{p}$, for $l=0$. In this same limit, all the remaining modes ($l \neq 0$) collapse in a single frequency at $\omega = \sqrt{3/2} \omega_{0} (1-\beta^{2})^{1/2} = \omega_{p} / \sqrt{2}$, characteristic of a cylindrical system. It is worth mentioning that for a spherical system, the same problem also yields a breathing mode at $\omega_{B} = \omega_{p}$ and, solving the Laplace equation in spherical coordinates \cite{hugo}, we would get $\omega = \omega_{p} \sqrt{l  / (2l+1)} = \omega_{0} \sqrt{3l / (2l+1)}$ which collapses into our second breathing mode in the limit $l \rightarrow \infty$, where we get $\omega = \omega_{p}/ \sqrt{2}$.

Notice that by reformulating the problem in terms of an effective dielectric function - see eq. (\ref{eq:eqp}) - we introduced the possibility for propagating modes in the boundless $z$ direction, and determined by radial boundary conditions, in eq. (\ref{eq:bound1}) and eq. (\ref{eq:bound2}). These modes have some similarities with the plasma waves considered in \cite{p1}, which take the form
\begin{equation}
\frac{\omega}{\omega_{p}} = \sqrt{1 + k_z^{2} \frac{u_{s}^{2}}{ \omega_{p}^{2}}} \, , 
\end{equation}
($u_{s}$ is the sound speed). In the low wavenumber limit, the dispersion relation of eq. (\ref{eq:omegap}), and figure (\ref{fig:dispersao}), takes the same form of these plasma waves, $\omega/\omega_{p} \simeq \sqrt{1/2 + k^{2}u_{l}^{2}/\omega_{p}^{2}}$, where $u_{l}$ is some constant for each $l$ solution, clarifying plasma like nature of these oscillations. In contrast, in the high wavenumber limit,  the radial boundary conditions  would considerably alter the dispersive relation, as indicated above. Such modes could be also be excited in bound extended systems, if the size in the $z$ direction is large compared with the typical wavelength $2 \pi/ k_z$. As mentioned before, a cylindrical geometry can experimentally be achieved by lowering the intensity of the laser cooling beams along the rotation axis, allowing for the cloud to expand and to take a cigar-like shape. Kinetic effects are not expected to play an important role here. In fact, a kinetic description is useful when considering the resonant exchange of energy between atoms moving with velocities near the phase velocity of phonons, $v_f = \omega / k$. This is not relevant for normal modes, which can be understood as phonons with infinite wavelength, $k \rightarrow 0$ or $v_f \rightarrow \infty$.

\section{Conclusion}

In this work we have examined the equilibrium profiles and the normal oscillating modes of rotating clouds of ultra cold atoms confined in a MOT.

The investigation relies on the assumption that a polytropic equation of state for the ultra cold gas is satisfied, which phenomenologically models a large class of MOT's, including the state-of-the-art ones, with a large number of atoms - in fact water-bag density profiles have been experimentally observed since the early stages of large MOTs \cite{mot2}. An experimental validation of this equation of state, which only depends on the polytropic exponent $\gamma$, can then be performed through the measurement of the density profiles of the atomic cloud, or by determining the spectrum of normal modes. Employing direct and absorption imaging techniques, this can be done in a very precise manner and would, for the first time, allow for the measurement of an equation of state of the ultra cold matter inside the trap, which constitutes one of the novelties of the present results.

On the other hand, we have shown that this simple polytropic model can account for the observed stable orbital modes, that up so far were thought to be intrinsically related with the cloud rotation \cite{mot2, 1ring, 2ring}. The profiles with orbital modes correspond to a special set of solutions with $\gamma = 3/2$. The measurement of the normal modes provides a nice way of validating these results.

The model employed here is valid for cylindrically symmetric clouds, which should be a good approximation in the equatorial plane of a spherical one. In order to determine the width of the satellite rings in spherical clouds, we would however need to develop a spherical model. In the general case, and specially for rotating clouds, the appropriate set of coordinates lies somewhere in between cylindrical and spherical. For this reason we are now trying to figure out how to employ an elliptical coordinate system that in the non-rotating case would reduce to the spherical geometry (zero eccentricity). This would be a more general and appropriate description of the cloud. These different aspects of the problem will be addressed in a future work.

Also related to these features, and following the annunciated new trend of studying astrophysical related phenomena in the laboratory, the results present here, as well as those in \cite{hugo}, allow for an investigation of the equilibrium and dynamical properties of the Lane-Emden equation, establishing another interesting connection between the communities of ultra-cold matter and astrophysics. In particular we can refer to the recent observation of a new class of variable stars \cite{astro}. These rapidly rotating main-sequence stars were not expected to pulsate, or to have any other physical characteristics that would lead to periodic luminosity variations, according to current theories. One of the hypothesis being considered is that fast rotation could alter the internal conditions of a star, enough to sustain stellar pulsations. But there is currently no stellar model that can predict whether pulsation can be sustained in very fast rotating stars. The ability to achieve conditions in laboratory, namely with ultra-cold atomic gases, with similar dynamics to stellar and astrophysical fluids, can prove being important to understand the properties of these objects. All the similarities between astrophysical and cold atoms systems comes from the concept of radiation pressure, which determines the dynamics of both systems. In astrophysics its the huge photon densities that leads to this radiation pressure. In cold atoms, the low densities are compensated by the high resonances leading to working regimes dictated by the radiation pressure - multiple scattering regime. In fact, it may be possible to tune the experimental parameters in order to achieve working conditions closer to those found in astrophysical scenarios.

\end{document}